\documentclass[english,aps,prper,reprint,showpacs,titlepage,longbibliography]{revtex4-2}
\title{Your Title Here}
\usepackage[utf8]{inputenc}
\usepackage{booktabs} % For better-looking tables
\usepackage{tabularx} % For flexible column widths
\usepackage{ragged2e} % For text alignment
\usepackage{caption} % For caption customization
\usepackage[T1]{fontenc}	
\usepackage{geometry}
\usepackage{times}
\usepackage{natbib}
\usepackage{array}

\usepackage{enumerate}
\usepackage{enumitem}
\usepackage{amsmath}
\usepackage{amssymb}
\usepackage{tikz}
\usepackage{graphicx}
\usepackage{multirow}
\usepackage{tcolorbox}
\usepackage{ragged2e}

\usepackage[utf8]{inputenc}

\usepackage[normalem]{ulem}
\usepackage{setspace} % Include the setspace package for line spacing control
\title{Help or Hype? Students' Engagement and Perception of Using AI to Solve Physics Problems}
\usepackage{hyperref}
\hypersetup{colorlinks=true,urlcolor=blue,citecolor=blue,linkcolor=blue} 
\begin{document}
\begin{titlepage}

\title{Help or Hype? Students' Engagement and Perception of Using AI to Solve Physics Problems}

 \author{Qurat-ul-Ann Mirza}
 \affiliation{Division of STEM, Rowan College at Burlington County, Mount Laurel, NJ 08054, U.S.A} 
 \affiliation{Depart. of Physics and Astronomy, Purdue University, West Lafayette, IN-47907, U.S.A.} 
  
 \author{N. Sanjay Rebello}
 \affiliation{Dept. of Physics and Astronomy / Dept. of Curriculum \& Instruction, Purdue University, West Lafayette, IN-47907, U.S.A.} 


\begin{abstract}
With the rise of large language models such as ChatGPT, interest has grown in understanding how these tools influence learning in STEM education, including physics. This study explores how students use ChatGPT during a physics problem-solving task embedded in a formal assessment. We analyzed patterns of AI usage and their relationship to student performance. Findings indicate that students who engaged with ChatGPT generally performed better than those who did not. Particularly, students who provided more complete and contextual prompts experienced greater benefits. Further, students who demonstrated overall positive gains collectively asked more conceptual questions than those who exhibited overall negative gains. However, the presence of incorrect AI-generated responses also underscores the importance of critically evaluating AI output. These results suggest that while AI can be a valuable aid in problem solving, its effectiveness depends significantly on how students use it, reinforcing the need to incorporate structured AI-literacy into STEM education.
  
\clearpage
\end{abstract}

\maketitle
\end{titlepage}
\maketitle

\section{Introduction}
Since the public release of generative artificial intelligence (AI) tools such as OpenAI’s ChatGPT, their integration into physics education has gained increasing attention. Numerous studies \cite{patero, liang, Baidoo, walter, mahligawati} have explored the educational potential of these tools, highlighting their ability to personalize learning, foster critical thinking, provide immediate feedback, and scaffold complex concepts. In particular, studies \cite{mahligawati, horchani, West, polverini} have shown that AI performs well on conceptual tasks and can enhance students’ understanding of foundational physics ideas. This is especially relevant in physics classrooms, where conceptual understanding is as critical as quantitative problem-solving.

More recent studies have shown that AI can successfully solve traditional physics problems \cite{horchani, CanAISolve, Mustofa} yet may still struggle with calculation errors, misinterpretation of physical scenarios, and inconsistent outputs \cite{CanAISolve}.

Given these findings, it is important to guide students toward using AI as a tool for learning rather than a shortcut to answers. In this study, we explore how students interact with AI when solving a modified but traditional kinematics problem, with the goal of understanding how AI can support meaningful engagement with physics content. The research questions we seek to answer are:

\begin{quote}
\textit{In what ways do students use ChatGPT to solve an unfamiliar physics problem? \\ What are students' perceptions of the use of AI in problem solving?}
\end{quote}
In Section~\ref {sec:Background}, we present other related work. The methodology is presented in Section~\ref{sec:Methods}. Results and discussion are presented in Section~\ref{sec:Findings}, leading into concluding remarks as well as a discussion of the study’s limitations and implications for future research in Section~\ref{sec:Conclusions}.  

\section{Background}
\label{sec:Background}
Recent research has explored the use of generative AI tools, such as ChatGPT, in physics education through a variety of designs, from exploratory surveys to structured classroom interventions.

Wang et al. surveyed 40 students on how they might use AI in problem solving, finding that 54\% would rely on it to directly solve problems and 50\% to save time, while most also valued its ability to explain concepts (82.5\%) and assist with calculations (65\%). However, participants did not solve problems during the study, instead describing their intended approaches \cite{wang}. Krupp et al. compared the performance of 39 STEM students using ChatGPT-3.5 Turbo or a traditional search engine, reporting significantly lower performance among ChatGPT users. This underperformance was linked to over-reliance on AI-generated responses—nearly half of which were incorrect—though the findings were limited by a small, non-random sample \cite{Krupp}.

Other studies point to potential benefits. Patero found a 27\% average improvement in student comprehension when using interactive AI tools, with participants citing increased engagement, motivation, and understanding, though the study lacked details on the tools and their implementation \cite{patero}. Bitzenbauer involved 53 secondary students in two quantum physics lessons using ChatGPT, where students critically evaluated AI-generated responses and developed conceptual survey items. While perceptions of ChatGPT improved, the study did not examine problem-solving performance or detailed usage patterns \cite{Biltzenburg}. Robledo-Rella and Toh implemented ChatGPT and Copilot in undergraduate physics courses, where students generated their own problems, attempted solutions, and then engaged with AI. Although they detected and corrected frequent AI errors (incorrect 95\% of the time), no significant learning gains were observed compared to traditional instruction, and no pre- or post-assessments of conceptual understanding were included \cite{Victor}.

Collectively, these studies illustrate varied approaches to AI use in physics, from surveys to structured classroom interventions. The present study offers a new perspective by focusing specifically on how students interact with AI in the context of problem-solving. By embedding AI use into a graded assessment and analyzing both prompt structure and student-AI interactions, we aim to provide deeper insight into how such tools can support—not replace—conceptual learning while solving problems in introductory university physics.

\section{Methods}
\label{sec:Methods}

\begin{table*}[!htbp]
\caption{The Pre-Quiz Problem was administered through an online assessment platform under monitored conditions. Problems 1 and 2 were included in the in-person assessment. These two problems were identical in structure, with the only difference being the launch angle of the firecracker.}
\begin{ruledtabular}
\begin{tabular}{p{0.14\linewidth} p{0.8\linewidth}p{0.8\linewidth}}
{\bf Problems}  & {\bf Description of the Problems} \\ 
\hline 
{Pre-Quiz}  & {A military helicopter on a training mission hovers 300 \(m\) above the ground. It accidentally launched an object horizontally at a speed of 60 \(m/s\). (You can ignore air resistance.)} \\
& \parbox{0.99\linewidth}{\begin{enumerate}[itemsep=-0.1 cm]
    \item How much time is required for the object to reach the ground? 
    \item How far does it travel horizontally while falling? 
    \item Find the magnitude of the final velocity vector.
    \item A military helicopter on a training mission hovers 300 \(m\) above the ground. It accidentally launched an object at an angle of \(25^\circ\) from the horizontal at a speed of 60 \(m/s\). What is the maximum height the object will reach with respect to the ground?
\end{enumerate}}\\
\hline 
{Problem 1}  & {A student sits atop a platform \textit{h} above the ground. He throws a large firecracker horizontally at a speed 10 \(m/s\). However, a wind blowing parallel to the ground gives the firecracker a constant horizontal acceleration with a magnitude of 7 \(m/s^2\), resulting in the firecracker landing directly under the student.} \\
& \parbox{0.99\linewidth}{\begin{enumerate} [itemsep=-0.1 cm]
    \item Determine the height at which the firecracker was thrown. 
    \item What is the maximum horizontal position of the firecracker?
    \item Draw the motion graphs of the firecracker in both directions. You should have a total of 6 graphs: (\(x\) vs. \(t\), \(v_x\) vs. \textit{t}, \(a_x\) vs. \(t\), \(y\) vs. \(t\), \(v_y\) vs. \(t\), \(a_y\) vs. \(t\))
\end{enumerate}}\\
\hline 
{Problem 2} & {A student sits atop a platform \textit{h} above the ground. He throws a large firecracker at an angle of \(30^\circ\) with a speed 10 \(m/s\). However, a wind blowing parallel to the ground gives the firecracker a constant horizontal acceleration with a magnitude of 7 \(m/s^2\), resulting in the firecracker reaching the ground directly under the student.}\\
& \parbox{0.99\linewidth}{\begin{itemize} [itemsep=-0.1 cm]
    \item Subparts same as Problem 1. 
\end{itemize}}\\
\end{tabular}
\end{ruledtabular}
\captionsetup{justification=raggedright,singlelinecheck=false} 
\label{tab:table1}
\end{table*}

This study involved undergraduate students enrolled in a calculus-based introductory physics course at a community college in the northeastern United States. The primary aim was to explore how students use AI in problem solving and how AI tools can guide them through the problem-solving process. To ensure students had sufficient contextual knowledge to critically evaluate AI-generated responses, the study was embedded within a graded assessment. The study consisted of two components: an online portion and an in-person session. Students were instructed to complete the online assessment (Pre-Quiz) prior to attending the in-person class session. To ensure academic integrity, LockDown Browser and Respondus Monitor were used on the online portion of the quiz. The primary objective was to assess students' baseline understanding of key kinematics concepts. 
The in-person quiz consisted of two problems centered on kinematics concepts, see Table \ref{tab:table1}. 
\begin{itemize}
    \item \textbf{Problem 1} was completed independently, without access to any AI tools. This ensured that the students' responses reflected their understanding and problem-solving approaches. Our purpose was to capture students' initial, unaided attempt at solving the problem.
    \item \textbf{Problem 2}: Students were encouraged to engage with ChatGPT to gain insight and develop an understanding of the problem. Simply copying and pasting of the question and having AI solve it was not permitted. This part was designed to evaluate how students utilize generative AI (ChatGPT) in the context of physics problem solving.
\end{itemize}
Both problems included constant horizontal acceleration, a concept that was not yet addressed in the course.

In addition to the assessment problems, students were asked a set of self-reflection questions to gather insights into their perceptions of using ChatGPT during the assessment and how their experience might influence their future use of the tool. These questions are outlined below:
\begin{itemize}
    \item What new insights or understanding did you gain about the problem from ChatGPT that you were previously unaware of? (Please provide specific examples.)
    \item How might your interactions with ChatGPT influence how you use it in the future?
\end{itemize}
Students' responses to problems, self-reflection questions, and the ChatGPT transcript were collected. The transcripts and reflections were qualitatively analyzed to identify emergent themes. 

%\begin{figure}[H]
    %\captionsetup{justification=raggedright,singlelinecheck=false} 
   % \centering
   % \fbox{\includegraphics[width=0.8\linewidth]{LearningGains.png}}
   % \caption{Distribution of Students by AI Usage and Learning Gains}
   % \label{fig:Learnig gains}
%\end{figure}

\section{Findings \&\ Discussion}
\label{sec:Findings}

We first discuss the results from our analysis of student performance on Problem 1 and Problem 2 and then describe the themes emergent from our analysis of students' written prompts and reflections.

\subsection{Student Performance}

A total of 49 students participated in the study in the Fall 2024 and Spring 2025 semesters. % Figure  \ref{fig:Learnig gains} presents the distribution of students based on their learning gains and AI usage.% 
Among the 39 students who used AI tools, 14 showed improved scores from Problem 1 to Problem 2, with an average increase of 0.43 points. In contrast, among the 10 students who chose not to use AI tools, only two students showed improved scores from Problem 1 to Problem 2, with an average decrease of 1.8 points. A Mann–Whitney U test yielded a p-value of 0.06, which is statistically significant at the $\alpha = 0.10$ level. These results suggest that students who engaged with AI tools showed higher learning gains than those who did not.

The correlation between the Pre-Quiz and Problem 1 score was +0.22 and the correlation between the Pre-Quiz and Problem 2 score was +0.21, neither of which were significant. This weak positive correlation suggests only a slight tendency for the students who performed well on the Pre-Quiz to do well again on the two Problems. The correlation between Problem 1 and Problem 2 is +0.74, indicating that students who performed well on Problem 1 also performed well on Problem 2.

\subsection{Students' Written Prompts \& Self-Reflections}

Analysis of student written prompts and reflections reveals several emergent themes. We discuss each theme below. \medskip

\begin{figure} [!htbp]
    \centering
    \captionsetup{justification=raggedright, singlelinecheck=false}
    \includegraphics[width=0.95\linewidth]{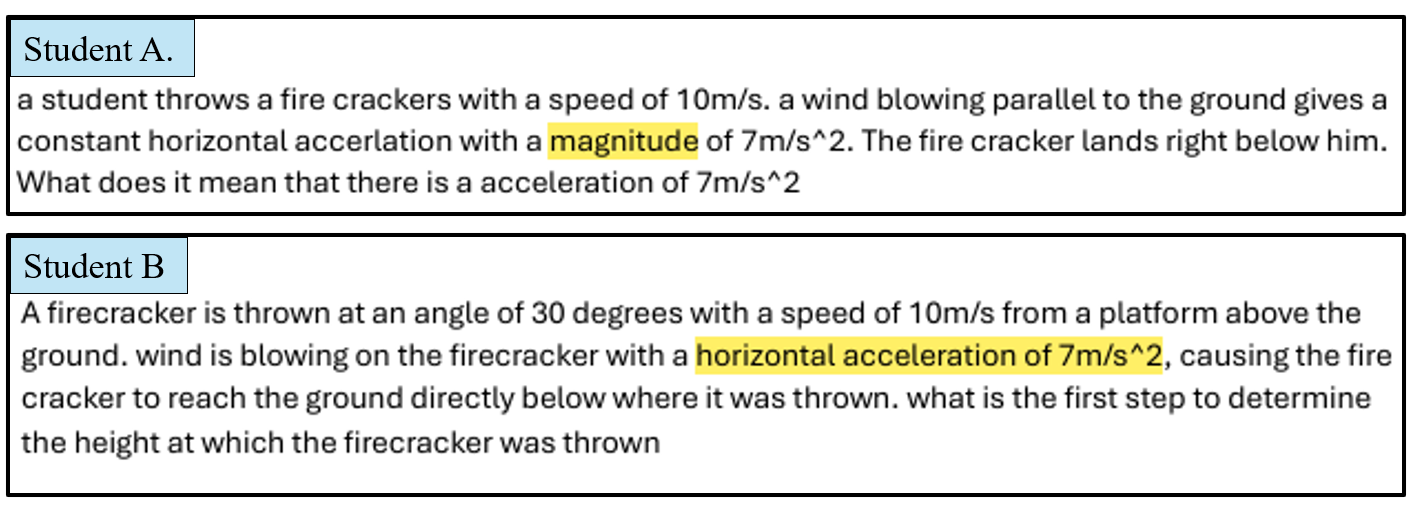}
    \caption{Comparison of students' initial prompts: Student A, who used a \textit{conceptual} prompt, saw a 10-point score increase. Student B, who used a \textit{procedural} prompt, saw an 8-point score decrease. Another difference between their prompts was the use of the word 'magnitude' in describing acceleration.}
    \label{promt1}
\end{figure}

\textbf{Conceptual vs. Procedural}:  Students who provided ChatGPT with prompts that included complete and detailed problem contexts, and asked \textit{conceptual} questions focused on idea or principle (e.g. "What does it \textit{mean} that ...?") generally received more accurate and helpful responses from ChatGPT. In contrast, students whose performance declined had often provided little to no information, or gave incomplete or partial descriptions of the problem. These students frequently omitted elements, including key variables such as acceleration and horizontal displacement, and collectively asked more \textit{procedural} questions focused on steps to solve the problem. ChatGPT’s responses for these students also tended to contain errors, most commonly involving incorrect assumptions about the direction of the acceleration vector and the inappropriate use of vertical motion equations to calculate time. Figure \ref{promt1} shows the prompts provided by the students with the maximum and minimum score gains. A notable difference was the completeness of the acceleration description: Student A included a full description specifying the magnitude of the acceleration, whereas Student B omitted the term “magnitude” (highlighted). Furthermore, Student A clarified the concept of acceleration, a topic with which they had previously struggled in Problem 1, while Student B (who had correctly solved Problem 1) focused on procedural prompts.

\begin{figure}
    \centering
    \captionsetup{justification=raggedright, singlelinecheck=false}
    \fbox{\includegraphics[width=0.9\linewidth]{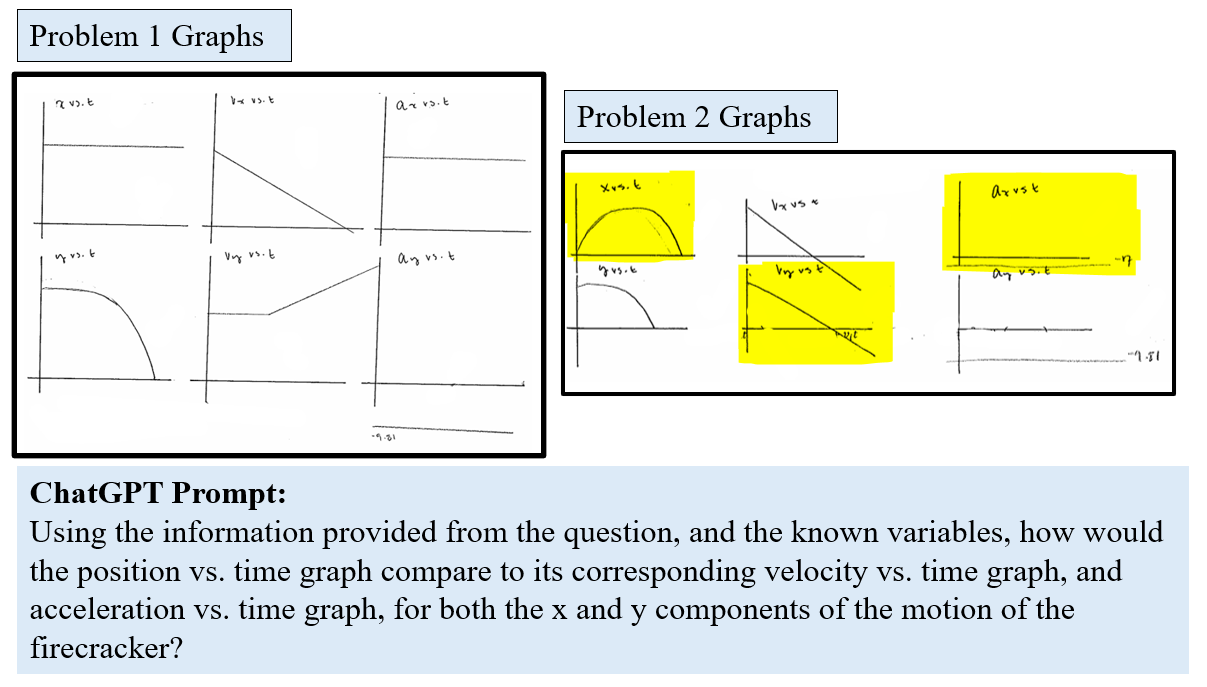}}
    \caption{Students' graphs for Problems 1 and 2, alongside their ChatGPT prompt. Notable differences include the \textit{x} vs. \textit{t} graphs (top left in both images), the \(v_y\) vs. \textit{t} graph (bottom center in both images), and \(a_x\) vs. \textit{t} graph (top right in both images).}
    \label{fig:graphexample}
\end{figure}

\textbf{Supporting vs. Verifying}:
Most students appeared to benefit from using AI for \textit{supporting} procedural steps or clarifying conceptual understandings. For example, ChatGPT was effective in assisting with the description of motion graphs. Figure \ref{fig:graphexample} presents an example of how students use AI to draw correct motion graphs in Problem 2. These interactions reveal the ways in which students engaged with AI to clarify concepts related to position, velocity, and acceleration.
\\
However, a smaller number of students used ChatGPT primarily for \textit{verifying} their work. Notably, a few students who did so also demonstrated critical thinking by identifying and correcting errors in the AI's responses. These findings suggest that while AI can be a valuable learning tool, its effectiveness depends heavily on the quality of input and the user's ability to engage critically with its output.

A similar pattern emerged from an analysis of students' reflections on their use of ChatGPT. Of the 49 students, 38 responded to the question about insights gained through using the tool. Fourteen students reported using ChatGPT primarily for clarification and conceptual understanding. Twenty students used ChatGPT to guide their problem-solving process, especially in decomposing velocity into components, identifying knowns and unknowns, and setting up equations. Many students appreciated the step-by-step structure that supported their understanding. Many students cited their improved understanding of horizontal acceleration, particularly why it is sometimes considered negative and how horizontal motion differs from vertical motion.\medskip

\textbf{Frustrations with AI}: Despite the aforementioned benefits of using ChatGPT cited by students, a few students reported dissatisfaction with using AI. Examples of students' complaints include 
\textit{"ChatGPT initially misinterpreted my input of the problem"}, \textit{"ChatGPT has a different way of writing our equations,"}, and \textit{"I feel like I got more confused trying to follow the AI as some steps in solving were sometimes skipped."} \\
One of the main reasons for the frustration is that the information provided by AI was somewhat different from what they had learned, which they found added to their confusion and cognitive load.\medskip

\textbf{Choosing \textit{not} to use AI}: While a majority of students expressed appreciation for ChatGPT’s step-by-step guidance and saw value in using it as a supplemental resource to reinforce concepts outside of class, 10 of the 49 students (about 20\%) chose not to use AI tools on Problem 2. Within this subset, five students scored 87\% or higher on Problem 1, whereas only two reached that threshold on Problem 2. The most commonly cited reasons for not using AI included confidence in their own understanding of the problem and unfamiliarity with the tool. Nonetheless, some of these students indicated they would be open to using generative AI in the future as a learning aid. Nevertheless, a portion of these students expressed openness to using AI in the future as a learning aid. 
This divide in perception was further evident when students were asked about their intentions for future use. Eight students indicated that they would not use the tool, often citing a preference for traditional learning methods or a desire to build understanding independently, rather than rely on an AI tool.

\section{Conclusions, Limitations, \&\ Future Work}
\label{sec:Conclusions}

This study contributes to the growing body of research on the use of AI in physics education by examining how students interact with generative AI tools during problem solving tasks. 

Our analysis of the students' performance shows that most students who used ChatGPT benefited from it, as evidenced by their improved performance with the use of ChatGPT. Several themes emerged from the analysis of students' written prompts and their self-reflections after the process.

First, we found a substantial variation in how students use ChatGPT to approach unfamiliar physics problems, including seeking procedural help, clarifying concepts, and verifying their answers.  Interestingly, students who used ChatGPT for \textit{conceptual} assistance e.g. asking for clarification on the meaning of information provided in the problem, seemed to benefit more from their use of ChatGPT than students who asked for \textit{procedural} help, e.g., asking for the next step in the solution. This provided anecdotal evidence of a greater improvement in scores for students who sought conceptual help from ChatGPT compared to those who sought procedural help.

Second, we also found that while most students used ChatGPT to \textit{support} their understanding of the conceptual and/or procedural aspects of a problem, a few used it to \textit{verify} their solutions to the problems. This latter group seemed to be more critical examining the output provided by ChatGPT.

Third, students' reflections highlighted both the advantages and limitations of using ChatGPT in this context. Many students appreciated its strategic guidance and used it to enhance their understanding of concepts, while others reported confusion due to differences in format, skipped steps, or overly complex explanations. Despite these challenges, most students regarded ChatGPT as a valuable supplemental tool, though a few expressed a preference for traditional approaches that foster independent understanding.

Finally, a small, but significant (about 20\%) of the participants chose \textit{not} to use ChatGPT because of their unfamiliarity with the tool or confidence in their understanding of the problem. While some students expressed that they would avoid using AI in the future, others viewed it as a potential tool to support their learning.

Overall, the results suggest that while tools like ChatGPT can support problem-solving and reinforce learning, their effectiveness depends significantly on how students formulate prompts, interpret AI responses, and apply critical thinking. As AI becomes more embedded in education, fostering AI literacy will be essential to help students engage as active, reflective learners rather than passive recipients of answers.

%\section{Limitations \&\ Future Work}
%\label{sec:Limitations}

This study has several limitations. First, the small sample size (N = 49) limits the generalizability of the findings. Second, data collection lacked details on which ChatGPT version was used, affecting response quality and consistency. The study also did not account for students’ prior experience with AI tools, which may have influenced usage and outcomes. Lastly, self-selection bias may be present, as students choose whether to use ChatGPT, potentially reflecting unmeasured factors like confidence or preferences for learning.

In future research, we will examine how training students in prompt engineering and the evaluation of AI-generated outputs influences their learning outcomes in physics problem solving.

\clearpage

% Make sure the citations are in the PRPER format. Download the bibtex file from the site where you found the paper. Google scholar may not be sufficient. See example below: 
%@article{sirnoorkar2023analyzing,
% author    = {Amogh Sirnoorkar and James T. Laverty},
% title     = {Analyzing Students’ Assumptions to Varying Degree of Prompting During Problem Solving},
% journal   = {Physics Education Research Proceedings},
% volume    = {302},
% year      = {2023},
% publisher = {American Association of Physics Teachers},
% doi       = {10.1119/perc.2023.pr.Sirnoorkar},
% url       = {https://doi.org/10.1119/perc.2023.pr.Sirnoorkar}
% }

\bibliography{references}
\bibliographystyle{ieeetr}
\end{document}